\documentclass{pasj00}

\begin{document}
\SetRunningHead{M.~S.~Tashiro et al.}{Swift and Suzaku Observation of the X-Ray Afterglow of the GRB~060105}
\Received{2006/07/30}
\Accepted{2006/09/04}

\title{{\it Swift} and {\it Suzaku}  Observations of the X-Ray Afterglow
 from the  GRB~060105}



%
 \author{%
   Makoto~S. \textsc{Tashiro}\altaffilmark{1},
   Keiichi \textsc{Abe}\altaffilmark{1},
   Lorella \textsc{Angelini}\altaffilmark{2},
   Scott \textsc{Barthelmy}\altaffilmark{2},\\
   Neil \textsc{Gehrels}\altaffilmark{2},
   Nobuyuki \textsc{Ishikawa}\altaffilmark{3},
   Louis~J. \textsc{Kaluzienski}\altaffilmark{2},\\ 
   Nobuyuki \textsc{Kawai}\altaffilmark{4},
   Richard~L. \textsc{Kelley}\altaffilmark{2},
   Kenzo \textsc{Kinugasa}\altaffilmark{5},
   Hironobu \textsc{Kodaira}\altaffilmark{6},\\ 
   Takayoshi \textsc{Kohmura}\altaffilmark{7},
  Kaori \textsc{Kubota}\altaffilmark{8},
   Yoshitomo \textsc{Maeda}\altaffilmark{9},
   Shouta \textsc{Maeno}\altaffilmark{10},\\ 
   Hiroshi \textsc{Murakami}\altaffilmark{9},
   Toshio \textsc{Murakami}\altaffilmark{6},
   Yujin~E. \textsc{Nakagawa}\altaffilmark{3},
   Kazuhiro \textsc{Nakazawa}\altaffilmark{9},\\ 
   John \textsc{Nousek}\altaffilmark{11},
   Shin'ya \textsc{Okuno}\altaffilmark{6},
   Kaori \textsc{Onda}\altaffilmark{1},
   James~N. \textsc{Reeves}\altaffilmark{2},\\
   George \textsc{Ricker}\altaffilmark{12},
   Goro \textsc{Sato}\altaffilmark{2,9},
   Eri \textsc{Sonoda}\altaffilmark{10},
   Motoko \textsc{Suzuki}\altaffilmark{13},\\ 
   Tadayuki \textsc{Takahashi}\altaffilmark{9},
   Toru \textsc{Tamagawa}\altaffilmark{13},
   Ken'ichi \textsc{Torii}\altaffilmark{14},
   Yoshihiro \textsc{Ueda}\altaffilmark{8},\\
   Yuji \textsc{Urata}\altaffilmark{1},
   Kazutaka \textsc{Yamaoka}\altaffilmark{3},
   Makoto \textsc{Yamauchi}\altaffilmark{10},
   Daisuke \textsc{Yonetoku}\altaffilmark{6},\\
   Atsumasa \textsc{Yoshida}\altaffilmark{3},
      and
   Satoru \textsc{Yoshinari}\altaffilmark{6},

}
 \altaffiltext{1}{Saitama University, Sakura, Saitama, 338-8570}
 \email{tashiro@phy.saitama-u.ac.jp}
 \altaffiltext{2}{NASA/Goddard Space Flight Center, Greenbelt, MD 20771, USA}
 \altaffiltext{3}{Aoyama Gakuin University, Sagamihara, Kanagawa, 229-8558}
 \altaffiltext{4}{Tokyo Institute of Technology, Ohokayama, Meguro, Tokyo, 152-8551} 
 \altaffiltext{5}{Gunma Astronomical Observatory, Takayama, Gunma, 377-0702}
 \altaffiltext{6}{Kanazawa University, Kanazawa, Ishikawa, 920-1192}
 \altaffiltext{7}{Kogakuin University, Hachioji, Tokyo, 192-0015}
 \altaffiltext{8}{Kyoto Univsersity, Sakyo-ku, Kyoto, 606-8502}
 \altaffiltext{9}{JAXA/Institute of Space and Astronautical Science, Sagamihara, Kanagawa, 229-8510}
 \altaffiltext{10}{University of Miyazaki, Gakuen-kibanadai, Miyazaki, 889-2192}
 \altaffiltext{11}{Pennsylvania State University, University Park, PA 16802, USA}
 \altaffiltext{12}{Massachusetts Institute of Technology, CSR, Cambridge, MA 02139 USA.}
 \altaffiltext{13}{RIKEN, Wako, Saitama, 351-0198} 
 \altaffiltext{14}{Osaka Univsersity, Toyonaka, Osaka, 560-0043} 
 

\maketitle

\begin{abstract}
Results are presented of early X-ray afterglow observations of GRB~060105 by {\it Swift} and {\it Suzaku}. The bright, long gamma-ray burst GRB~060105 triggered the {\it Swift} Burst Alert Telescope (BAT) at 06:49:28 on 5 January 2006.
The {\it Swift} X-Ray Telescope (XRT) started pointed observations 87 s after the trigger ($T_0$).  The {\it Suzaku} team commenced a pre-planned target of opportunity observation at 19 ks (5.3 hr) after the {\it Swift} trigger as the first{\it Suzaku} attempt at a rapid response to a new gamma-ray burst. 
The X-ray flux faded during the observations from $6.8 \times 10^{-9}$ erg~s$^{-1}$cm$^{-2}$ (at $T_0 + 87$ s with the {\it Swift}/XRT) to $1.5 \times 10^{-13}$ erg~s$^{-1}$cm$^{-2}$ (at $T_0 + 94$--$101$ ks with the {\it Suzaku} X-ray Imaging Spectrometer (XIS)) in the 2--10 keV energy band.
The afterglow exhibited four phases of decay consisting of different decay indices during the {\it Swift}/XRT and {\it Suzaku}/XIS observations.
The inferred flux and spectral shapes with both instruments are fully consistent within the statistics, 
although there are hardening in the first shallow decay phase
and softening in very steep decay phases in the end of the observations.
Following the prompt emission and successive very steep decay, a shallow decay was observed from $T_0+187$ s to $T_0+1287$ s.
After an observation gap during $T_0 +(1.5$--$3)$ ks, an extremely early steep decay was observed in $T_0+(4$--$30)$ ks. 
The lightcurve flattened again at $T_0+30$ ks, and another steep decay followed from $T_0+50$ ks to the end of observations. 
Both steep decays exhibited decay indices $\alpha \sim$ 2.3 -- 2.4.
This lightcurve behavior can be explained as a steep decay resulting from side expansion of a jet, while the two flattening portions suggest energy injection and refreshed shock passage, respectively.
This very early break, if it is a jet break, is the earliest case among X-ray afterglow observations, suggesting a very narrow jet whose opening angle is well below $1^\circ$.
The unique {\it Suzaku}/XIS data allow us to set very tight upper limits on line emission or absorption in this GRB. 
For the reported pseudo-redshift of $z=4.0\pm1.3$ the upper limit on the iron line equivalent width is 50 eV. 
\end{abstract}

\section{Introduction}\label{sec:intro} 
Gamma-ray bursts (GRB) are thought to be caused by a sudden release of a large amount of energy ($\sim 10^{51}$ erg for long GRBs), on the scale of a single star.
The prompt gamma-ray emission is attributed to internal shocks in a highly relativistic outflow (\cite{rees94}; \cite{sari97}) with a Lorentz factor of over 100 (see \cite{piran99}, and references therein). After this early phase, relativistic ejecta sweep up a sufficient amount of external medium and are decelerated to cause external shocks.
A highly relativistic forward shock propagates into the circumstellar medium to produce the afterglow which can be observed across a wide energy band.

Before {\it Swift} (\cite{gehrels04}),
X-ray afterglow observations typically started several hours after the burst onset, and they showed a smooth single power-law decay $\sim t^{-1}$. In contrast, optical afterglow lightcurves often showed an achromatic steepening to $\sim t^{-2}$, which is attributed to the sideways expansion of a narrow jet (\cite{rhoads99}; \cite{sph99}). The advent of {\it Swift} has opened the earlier time window from $10^2$ -- $10^4$ s with the onboard X-Ray Telescope (XRT; \cite{burrows05a}). The {\it early} --- the first few hours --- afterglow evolution is a
probe not only of the early radiative energy losses from the external shock and energy injection, but also of the chemical abundance and density profile of the circumburst medium.

Because {\it Swift} observes X-ray afterglows much more rapidly and so has revealed a fundamental discrepancy from the simple power-law decay (e.g. \cite{nousek06}). These data show the limitations of a simple ``fire-ball model'' assuming a delta function-like ``sudden'' energy injection with self-similar evolution, and implies hidden physical processes.
Zhang et al. (2006) interpreted the Nousek et al. (2006) canonical afterglow lightcurve as having four phases: (i) very steep decay ($F_\nu (t) \propto t^{-\alpha}; \alpha \sim 3- 5$); (ii) (very) shallow decay ($\alpha \sim 0.5$); and (iii) somewhat steeper (or normal) decay ($\alpha \sim 1)$, where $F_\nu (t)$, $t$, and $\alpha$ are the X-ray flux, time from the burst, and decay index, respectively. In addition to those, (iv) jetlike (steep) decay ($\alpha \sim 2$); and temporal X-ray flares are observed in many GRBs (e.g. \cite{burrows05b}).

Each transition suggests an evolution of the physical state, such as a start or end of energy injection, re-acceleration by a reverse or forward shock, deceleration of a structured or uniform jet, and/or spatial distribution of physical parameters. However, no significant spectral transition has been observed. Precise 
spectroscopy, utilizing a high sensitivity X-ray telescope, is the key to determining any spectral evolution at the phase transitions, which will enable us to understand what physics controls phase transitions in GRB jets. Observations with high sensitivity and high energy resolution also examine the question of line features reported by some authors, but which
have not been confirmed by statistically strong detections
(\cite{piro99}; \cite{yoshida01}; \cite{reeves02}; \cite{butler03};
\cite{watson03}; and \cite{sako05}).
X-ray spectral data also provide limits on the environment in the very near vicinity around the gamma-ray burst and its progenitor.

We utilized the fifth Japanese X-ray observatory, {\it Suzaku} (\cite{suzaku06}), to conduct a follow-up observation of GRB 060105. Aiming to observe the X-ray afterglow in the fastest
possible time, we organized a Target of Opportunity (ToO) team to watch and respond quickly to {\it Swift} notices with {\it Suzaku} during the initial performance verification phase --- the Science Working Group time.

\section{Observations} 
\subsection{Early phase observatinos}\label{sub:swift}
The {\it Swift} Burst Alert Telescope (BAT; \cite{barthelmy05}) was triggered by
GRB~060105 at 06:49:28 UT on January 5, 2006 (hereafter $T_0$; \cite{gcn4429}; trigger number 175942) and located it at $(\alpha, \delta)$ = (\timeform{19h49m54s},\timeform{+46D21'45"}) (J2000).
{\it HETE-2} also noticed the GRB at 06:50:14 via the GRB Coordinate Network (GCN). 
With Konus-{\it Wind}, Golenetskii et al. (2006) reported the 20 keV to 2 MeV fluence of $(7.86^{+0.19}_{-0.37}) \times 10^{-5}$ erg~cm$^{-2}$
with a power-law with an exponential cutoff model spectrum, whose peak energy is
$E_p = 424_{-22}^{+25}$ keV.
The {\it Swift}/BAT lightcurve of the prompt emission is shown in figure~\ref{fig:batlc}, in which we see a precursor and two major peaks. A similar three peaked lightcurve was also confirmed with the {\it Suzaku} Wideband All-sky Monitor (WAM; \cite{yamaoka06}) in a harder (100 keV to 1 MeV) band (\cite{ohno06}).

The {\it Swift}/XRT began observations at 06:50:55.3 UT ($T_0+87$ s), and discovered a fading uncataloged X-ray source within the BAT error circle.
The {\it Swift}/XRT continued the observation in windowed timing (WT) mode
from $T_0+187$ s to $T_0+1,287$ s, automatically changing to photon counting (PC) mode  from  $T_0+1,431$ s to $T_0+24$ hr.
Although the UV/Optical Telescope (UVOT; \cite{roming05}) took a finding chart exposure of 200 s with the V filter starting 91 s after the BAT trigger, no afterglow candidate was observed. Neither optical nor radio afterglow has been reported with ground observations
(e.g. \cite{gcn4430}; \cite{gcn4431}; \cite{gcn4433}; \cite{gcn4436};
\cite{gcn4437}; \cite{gcn4438}; \cite{gcn4440}; \cite{gcn4441}; \cite{gcn4454}).

\subsection{Follow up observation with Suzaku}
The {\it Suzaku} GRB Target of Opportunity (ToO) team received the {\it Swift} notice and quick look result via the GCN at 87 s and 67 min after the trigger, respectively (\cite{gcn4429}).
The reported position satisfied the {\it Suzaku} solar angle constraint
($65^\circ < \theta_{\rm sun-z} < 110^\circ$), and the initial X-ray afterglow flux was high enough ($F_{\rm 2-10 keV} > 10^{-11}$ erg~cm$^{-2}$s$^{-1}$) to project that {\it Suzaku} instruments could measure the X-ray spectrum hours after the trigger. 
Upon receiving these notices the {\it Suzaku} steering committee approved the ToO observation.
As the timing of this decision was 1.25 hrs before the first of five contact orbits, the specially organized team prepared commands to upload to {\it Suzaku} on
the third contact orbit. 
Observations started at 12:10 ($T_0+5.3$ hr) and ended at 12:00 on the next day ($T_0+29.2$ hr). 
The {\it Suzaku} X-ray Imaging Spectrometer (XIS; \cite{xis06}), installed at the focal plane of the X-Ray Telescopes (XRT; \cite{xrt06}), immediately detected an uncatalogued fading source, thus confirming the detection of the X-ray afterglow of GRB060105.
The quick-look and refined analysis results were reported by Mitsuda et al. (2006), and Nakazawa et al. (2006), respectively.

\section{Data Analysis and Results}

\subsection{Early phase observation with {\it Swift}/BAT}
\begin{figure}
  \begin{center}
    \FigureFile(100mm,100mm){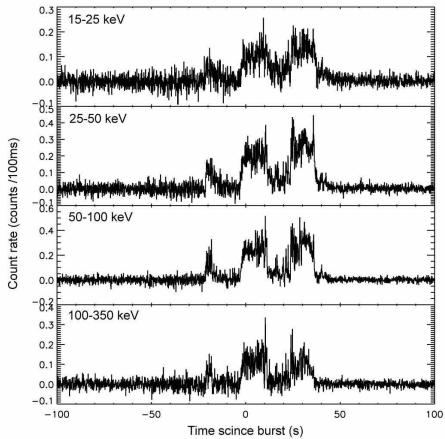}
  \end{center}
  \caption{The BAT lightcurve of the GRB~060105 prompt emission. All 15 -- 350 keV data are added, but subtracted both X-ray and non-X-ray background with
  the mask-weighted method.}\label{fig:batlc}
\end{figure}

The BAT data were analyzed using the standard BAT analysis software within {\tt heasoft 6.0.6}. Hereafter the quoted errors are at the 90 \% confidence level value. 
Mask-tagged BAT lightcurves for the 15 -- 150 keV band at 0.1 s time resolution are shown in figure~\ref{fig:batlc}. 
The 15 -- 150 keV fluence was $(1.82 \pm 0.04) \times 10^{-5}$ erg~cm$^{-2}$, as reported by Markwardt et al. (2006), and the $T_{90}$ (the time interval over which the flux is within 90\% of the peak burst photon fluence), calculated from the lightcurves was $54.4\pm1.4$ s.

\subsection{Afterglow observations with {\it Swift} and {\it Suzaku}}
\label{sub:afterglow}
{\it Swift}/XRT data were first processed by the {\it Swift} Data Center at NASA/GSFC into Level 1 products. We further processed them using the {\tt XRTDAS} software package to produce a final cleaned event list. In particular, we processed the retrieved data with {\tt xrtpipeline} and calibration data distributed with the {\tt heasoft 6.0.6}. We used only grade 0--12 and 0--2 for PC and WT mode data, respectively (according to {\it Swift} nomenclature; \cite{burrows05a}).

An X-ray source was confirmed at $(\alpha, \delta)$ = (\timeform{19h50m00s.6},\timeform{+46D20'58".3}) with an estimated uncertainty of \timeform{3".5}, which was exactly the same position reported by Godet et al. (2006). 
The data before $T_0+10$ks were affected by pile-up in the PC-mode observation. Our position takes into account correction for misalignment between telescope and the satellite optical axis.


\begin{figure}
  \begin{center}
    \FigureFile(100mm,100mm){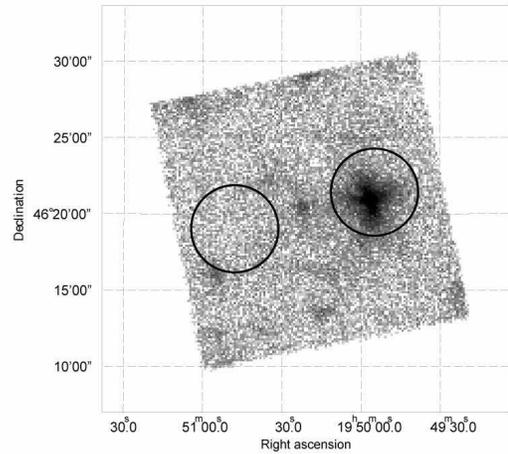}
  \end{center}
  \caption{The synthesized X-ray image obtained with {\it Suzaku}/XISs.
  	X-ray photons above 5.5 keV are removed to reduce calibration 
  	source image.
           The west and east circles represent the data accumulation regions
           from the source and background, respectively. 
           }\label{fig:suzaku_image}
\end{figure}


{\it Suzaku} data were extracted using the revision 0.7 pipeline process products.
We used data from the BI chip (XIS1) and FI chips (XIS0, XIS2 and XIS3), except for the first 24 minutes of data because {\it Suzaku} had not yet reached a stable attitude.
 The net exposure time was 35.4 ks.
Figure~\ref{fig:suzaku_image} shows the synthesized image of the four XIS fields of view.
The brightest west source is located at the position,
$(\alpha, \delta)$ = (\timeform{19h49m56s},\timeform{+46D20'56"}).
It is consistent with that reported by {\it Swift}/XRT within the accuracy at this stage of calibration (\cite{xrt06}).
We accumulated events within \timeform{2'.86} from the observed source center, while background events are also taken within the same radius but masked to exclude the contaminating point source and set at the axisymetric position on the optical axis of the {\it Suzaku}/XRT. The two circular regions are shown in figure~\ref{fig:suzaku_image}.


\begin{figure}
  \begin{center}
    \FigureFile(70mm,70mm){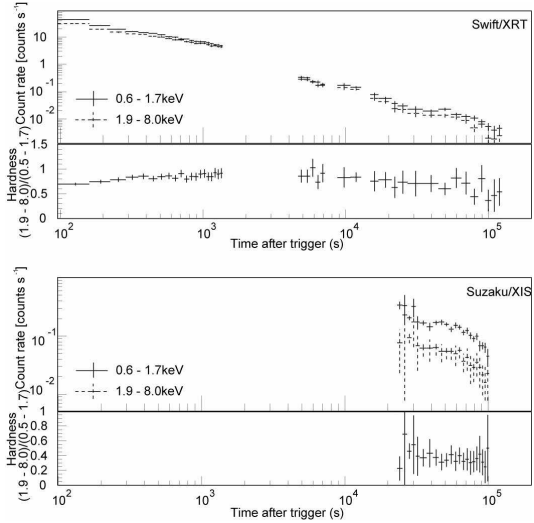}
  \end{center}
  \caption{The {\it Swift}/XRT (upper) and {\it Suzaku}/XIS (lower)
  background subtracted count rates in the soft (0.5 -- 1.7 keV; cross with solid lines)  
  and the hard (1.9 -- 8 keV; cross with dashed lines) bands
  and their hardness ratios ((1.9 -- 8 keV)/(0.5 -- 1.7 keV)) 
  are presented. The quoted errors are the 90 \% confidence level.
  Pile-up correction for data during $T_0+(15-20)$ ks has been applied.
  }\label{fig:hardness}
\end{figure}


{\it Swift}/XRT and {\it Suzaku}/XIS lightcurves in the soft (0.5 -- 1.7 keV) and the hard (1.9 -- 8.0 keV) bands are jointly shown in figure~\ref{fig:hardness} with the derived hardness ratios.
It is obvious that these lightcurves are not represented by a simple power-law like decay, but exhibit structures around $T_0+3.5$ ks and $T_0+40$ ks.
Although it is not significant, the ratio indicates a slightly hard spectrum before $T_0+10$ks. 
Apparent small variations in hardness near $T_0+(15$--$20)$ ks
may be due to pile-up problems in the PC mode data.
The possible spectral variation will be tested after the correction below.
After $T_0+20$ ks, the XIS hardness keeps constant despite the significant variation of decay rate, although the XRT hardness shows a sign of variation.

\begin{longtable}{*{1}{c}}
 \caption{The best fit parameters for the time averaged spectra}\label{tab:spec}
 \begin{tabular}{cccc}
 \hline \hline 
  duration$^\dagger$  $(T_0+$ (s))& $N_{\rm H} (10^{22}$ cm$^{-2}$) & energy index ($\beta$)  & $\chi_\nu^2$/dof\\
 \hline 
  (a) 187 --- 1,287     &  $0.40 \pm 0.02$    & $1.15 \pm 0.03$  & 1.12/733\\
  (b) 4,800 --- 12,000  &  $0.37 \pm 0.08$    & $1.07 \pm 0.17$  & 0.915/106\\
  (c) 20,520 --- 120,000 & $0.30 \pm 0.03$    & $1.14 \pm 0.05$  & 1.02/692   \\
  (c-1) PC only &  $0.30 \pm 0.06$ &        $1.09 \pm 0.13$  & 0.731/192 \\
  (c-2) XIS only &  $0.28 \pm 0.05$ &       $1.15 \pm 0.10$  & 1.13/499 \\
 \hline 
   \end{tabular} 
\\
$\dagger$: The evaluations are carried out for the time averaged spectra from (a) {\it Swift}/XRT-WT mode data;\\
(b) pile-up corrected {\it Swift}/XRT-PC mode data; (c) the overlapped region of {\it Swift}/XRT-PC mode.
\endfoot
\end{longtable}



\begin{figure}
  \begin{center}
    \FigureFile(70mm,70mm){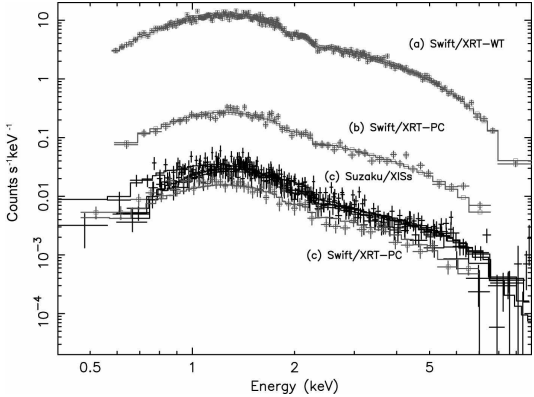}
  \end{center}
  \caption{The averaged XRT-XIS spectra for the time regions:
  	(a) from data obtained with the {\it Swift}/XRT-WT mode;
  	(b) with the {\it Swift}/XRT-PC mode; and
  	(c) with both the {\it Swift}/XRT-PC mode and
  	the {\it Suzaku}/XISs. 
  	The {\it Swift}/XRT data are represented with 
  	dark gray square marks, while {\it Suzaku}/XIS data are with black 
  	crosses.
  	The time durations, 
  	the obtained best-fit parameters
  	and the reduced chi-squares are shown in table~\ref{tab:spec}}
  	\label{fig:spectra}
\end{figure}

\begin{figure}
  \begin{center}
    \FigureFile(70mm,70mm){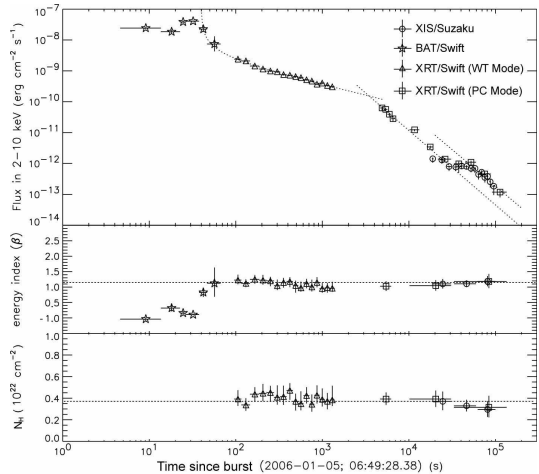}
  \end{center}
  \caption{The time history of the derived flux (upper panel), spectral index
  (middle panel) and absorption column density (lower panel) 
  from the {\it Swift} and {\it Suzaku} observations. 
  The indicated dotted lines represent the best fit functions
  for the decay lightcurves (see text and table~\ref{tab:phases}).
  Notice that we extrapolate and show the {\it Swift}/BAT 15 -- 350 keV spectra to evaluate the expected flux in 2 -- 10 keV. 
   The dashed lines in the middle and lower panels indicates the averaged
   values measured with the {\it Swift}/XRT and {\it Suzaku}/XIS 
   (\S~ref{sub:afterglow}).   
    }\label{fig:spectral_variability}
\end{figure}

In order to test for possible spectral variation in the afterglow, we
examined the averaged spectra over the following time regions:
(a) during the {\it Swift}/XRT-WT mode; (b) {\it Swift}/XRT-PC mode before $T_0+30$ks; and (c) the duration of the {\it Suzaku}/XIS observation.
We corrected the data in (b) suffering from pile-up for spectral analysis, according to Nousek et al. (2006). 
The consistency between {\it Swift}/XRT and {\it Suzaku}/XIS spectra is confirmed by independent fitting for the time region (c), as shown in table~\ref{tab:spec}, compared to a simultaneous fit for the {\it Swift}/{\it Suzaku} joint observation (c). These spectra are shown in figure~\ref{fig:spectra} and the best fit values are summarized in table~\ref{tab:spec}.
We see each spectrum is very well reproduced with an absorbed power-law model,
and each spectrum is consistent each other.
We evaluated averaged {\it Swift}/XRT and {\it Suzaku}/XIS spectra simultaneously. The derived best-fit values of energy index and absorption column density were $\beta = 1.15 \pm 0.03$ and  $N_{\rm H} = (3.74 \pm 0.03) \times 10^{21}$ cm$^{-2}$ , respectively.
Since the derived column density is comparable but significantly larger than that estimated for the measured Galactic HI values on the line of sight ($N_{\rm H}^{\rm Gal} = (1.56 - 1.59) \times 10^{21}$cm$^{-2}$; \cite{nh05}), the rest of the absorption is to be attributed to the host galaxy.
Although the spectral index ($\beta$) from the second spectrum of XRT-PC mode observation in (b) prefers a flatter slope, it is consistent within errors to those in (a) or (c).
In order to look through the spectral evolution during the {\it Swift/Suzaku} observations in short time scales, we performed time resolved spectral fitting and derived the results in figure~\ref{fig:spectral_variability} with the derived 2 -- 10 keV X-ray fluxes. 
We see a trend of spectral hardening during the {\it Swift}/XRT-WT mode observation ((a) in table~\ref{tab:spec})and the spectral index smoothly connects to that of the beginning of the PC mode (b). In addition to that, {\it Swift}/XRT-PC mode data and {\it Suzaku}/XIS data consistently infers rather less column density toward the end of observation (c). 
These trends are consistent with the signs we saw in figure~\ref{fig:hardness}.


\begin{longtable}{*{1}{c}}
 \caption{Phases of GRB~060105 X-ray Early Afterglow and Their Decay Indices}
 \label{tab:phases}
    \begin{tabular}{ccccc}
\hline \hline 
phase &  name          & duration ($T_0 +$ (s))     & decay index ($\alpha$)$^\dagger$ & $\chi^2_\nu$/d.o.f\\
\hline 
1  & first flat decay & 40 --- 1,287           & $0.691 \pm 0.019$
       & 1.05/15\\
2  & first steep decay & 4,800 --- 30,000         & $2.41 \pm0.14^\ddagger$
       & 3.15/8 \\ 
3  & second flat decay & 30,000 --- 50,000  & --- 
       & ---\\
4  & second steep decay & 50,000 --- 120,000 & $2.36 \pm 0.25$ 
       & 0.448/9 \\
\hline 
    \end{tabular}
 \\
$\dagger$: The errors shown in this column represent 68\% confidence levels.
Due to the limitation of statistics,\\ decay index of 
the phase 3 was  not derived. Details are described in \S~\ref{sec:discussion}.
\\
$\ddagger$: The derived break time is 0.04 day.
\endfoot
\end{longtable}

Above 10 keV, the extrapolated X-ray flux from the XIS spectra corresponds to $\sim 10$~\% of the cosmic X-ray background, which corresponds to half of the instrumental background count rate of the {\it Suzaku} Hard X-ray Detector (HXD; \cite{takahasi06}; \cite{kokubun06}).
Since the current background model has a few percent uncertainties, we cannot quote strict results at this calibration stage.
Conclusions on the HXD results must await improved calibrations. 

\subsection{Search for X-ray emission line features}
We performed a detailed inspection to search for the possible spectral features, using the {\it Suzaku}/XIS and {\it Swift}/XRT. 
However, we found no signature of any line feature during the summed spectra which spanned the entire observation time from 87 s to $1.2 \times 10^{5}$ s after the trigger. Even with the time-resolved spectral analyses as studied in previous work (see \S~\ref{sec:intro}), we found no emission line features. 
In summary we present our upper-limits on equivalent width for the neutral iron emission line assuming several redshifts in table~\ref{tab:lines}.


\begin{longtable}{*{1}{c}}
  \caption{Upper limits of iron lines equivalent width (EW)
   at possible redshifts}\label{tab:lines}
    \begin{tabular}{rcccc}
\hline \hline 
E(keV)   &  (a)XRT-WT (eV)   &    (b)XRT-PC (eV)  &  (c) 4XIS + PC (eV)  &  redshift\\
\hline 
   6.40  &   $< 25.3$ &     $< 530.$  &   $< 213.$       &   0 \\
   3.20  &   $< 44.2$ &     $< 71.0$  &   $< 58.4$       &   1\\
   2.13  &   $< 48.0$ &     $< 88.8$  &   $< 85.5$       &   2\\
   1.60  &   $< 32.0$ &     $< 64.4$  &   $< 11.0$       &   3\\
   1.28  &   $< 16.4$ &     $< 245$   &   $< 14.4$       &   4\\
   1.07  &   $< 20.6$ &     $< 79.2$  &   $< 50.6$       &   5\\
\hline 
    \end{tabular}
\\
     Employed time regions and data are the same as that used in table~\ref{tab:spec}. All the upper limits are calculated for 90\% error.
\endfoot

\end{longtable}


\section{Discussion}\label{sec:discussion}
\subsection{Decay time constants in different phases}
We observed the X-ray afterglow of the GRB~060105 with {\it Swift} and {\it Suzaku} through $T_0+87$ s to $T_0+29$ hr almost continuously. The 0.5 to 8 keV spectra, obtained with {\it Swift}/XRT and {\it Suzaku}/XIS, were well described with an absorbed power-law model.
On the other hand, the observed spectral index ($\beta \sim 1.1$) suggests that the X-ray bands are above the synchrotron cooling frequency.
If so, the spectral index of electron energy distribution is $p \sim  2.2$, where electron number density is $n \propto E^{-p}$.

In the early X-ray afterglow lightcurve (figure~\ref{fig:spectral_variability}), we identified four phase of decays as shown in table~\ref{tab:phases}. Each phase decay was successfully fitted with the function of $F(t) \propto (t - t_{\rm offset})^{-\alpha}$, where $F(t)$, $t$ and $t_{\rm offset}$ are the derived X-ray flux, time from the {\it Swift} trigger ($T_0$), and time offset from the trigger, respectively. First we derived the best fit value of $t_{\rm offset} = 39.7 \pm 0.9$ s and $\alpha_1 = 0.691 \pm 0.019$, from the phase 1 data. Then we fitted other two phase (2 and 4) decay indices using the previously derived $t_{\rm offset} $ as a fixed parameter (table~\ref{tab:phases}). The slope of decay in phase 1 seems to be consistent with those of both ``shallow (or somewhat steeper) decay'' and ``classical (or normal) decay'' (e.g. \cite{nousek06}). 

As for the phase 2 decay, the obtained decay index ($\alpha_2 = 2.41 \pm 0.14$) is too steep to regard it as the classical decay. It is consistent with the post-jet-break decay, although the rather large reduced $\chi^2_\nu$ value may suggest that it includes possible small X-ray flares. 
We saw no discontinuity in spectral index between phase 1 and 2. The decay without spectral change supports geometric changes in emission region like jet-break.

According to Sari, Piran, \& Halpern (1999), the expected jetlike decay index is $\alpha_{\rm jet} = 2 \beta$, as long as the observed band is above the synchrotron cooling frequency. The observed spectral index ($\beta \sim 1.1$) implies that the decay indices ($\alpha_2$ and $\alpha_4$) in phase 2 and 4 are consistent with those expected for a jetlike decay. In addition, the fact that the spectral index remains the same after the flattening (in the phase 3) supports the hypothesis that the cooling time scale is shorter than the time scale of the observation. On the other hand, the electron energy index, $p \sim  2.2$ --- estimated above, requires the decay index of 1.15 in the ``classical decay regime'', which is much larger than that observed in phase 1. Therefore we could regard phase 1 as the ``shallow decay'', and we suspect the expected ``classical decay'' could be during the data gap between phase 1 and 2. 
We also note that the suggested spectral hardening in phase 1 infers energy injection expected in the ``shallow decay'' phase,
If this is the case, the jet break should occur between the cross point of extrapolations of the decays of phases 1 and 2 and the beginning of the observed phase 2. This requires the jet break time of $T_0+(3.5-4.0)$ ks ($\sim 0.04$ day). This is the earliest jet break so far reported from the X-ray afterglow observations.

After the flat decay phase 3, we see another steep decay in phase 4 ($T_0+(50-120)$ ks). 
The evaluated decay index of $\alpha_4 = 2.36 \pm 0.25$ is consistent with the decay index in phase 2.
The jet-like steep decay suggests the decay was also caused by side expansion of the emission region.
We also note that X-ray spectra observed in phases 3 and 4 infer gradual softening in the steep decay phases (figure~\ref{fig:spectral_variability}). This may implies a change of column density along the emission regions, since it does not require a steeper slope but a lower column density  (table~\ref{tab:spec} (c) in comparison with those of (a)). However, we cannot reject gradual softening of the continuum. We need detailed examinations of the energy response and backgrounds of instruments before conclusion, since the fitting parameters of energy index and column density couples tightly.

\subsection{Estimation of the jet opening angle and the kinetic energy}
Under a simple assumption of a circumburst density medium of constant number density $n$, and a uniform jet emitting a fraction $\eta_\gamma$ of its kinetic energy in the prompt gamma-ray phase, and the afterglow would show the jet break when its bulk Lorentz factor $\Gamma$ is decelerated to the order of $\Gamma \sim 1/\theta$. The $\theta$ is estimated from equation (1) of Ghirlanda, Ghisellini \& Lazzani (2004) or equation (1) of Sari, Piran \& Halpern (1999). For the case of GRB~060105, the jet opening angle $\theta$ is estimated as,
\begin{equation}
\theta = 0.026 \left( \frac{t_{jet,d}}{0.04} \right)^{3/8}
	\left(\frac{5}{1+z} \right)^{3/8}
	\left( \frac{n \eta_\gamma}{E_{\gamma, iso, 52}} \right)^{1/8},
\end{equation}
where $z$ is the redshift,
$t_{jet,d}$ is the break time in days, and $E_{\gamma, iso}$ is the
energy in gamma-rays calculated assuming that the emission is
{\it isotoropic}.

Although the host galaxy is not detected (\cite{kann06}),
if we adopt the reported 20 keV to 2 MeV fluence (\cite{gcn4439}; see also \S~\ref{sub:swift}) and the estimated ``pseudo'' redshift $pz = 4.0\pm1.3$ (\cite{gcn4442}), $E_{\gamma, iso}$ could be $2.5 \times 10^{54}$ erg. Here we used the Hubble constant $H_0 = 71$ km~s$^{-1}$Mpc$^{-1}$, $\Omega_{\rm M} = 0.27$, and $\Omega_{\rm vac} = 0.73$ to estimate luminosity distance. In this case, the jet opening angle ($\theta$) should be $0.012^{+0.003}_{-0.002}$ rad and the {\it collimation-corrected} energy $E_\gamma (\equiv (1- \cos \theta) E_{\gamma, iso} )$ is $(1.8^{+0.3}_{-0.4}) \times 10^{50}$ erg. Here we take $n=3$~cm$^{-3}$ and $\eta_{\gamma} = 0.2$ according to Ghirlanda, Ghisellini, \& Lazzati (2004).

The estimated $E_\gamma$, however, is far below the expected 
$E_\gamma$-$E_p$ (Ghirlanda) relation (\cite{ggl04}). 
Above, we evaluated $E_{\gamma, iso}$ assuming the $E_p$-$E_{iso}$ relation (\cite{amati02}) using $E_p$ given by Golenetskii et al. (2006). Then, regarding the phase 1 -- 2 transition as the jet break, we estimated the jet opening angle ($\theta$) and resultant $E_{\gamma}$, according to Ghirlanda, Ghisellini \& Lazzani (2004). As far as we accept the $E_p$-$E_{iso}$ relation, this discrepancy implies two possibilities: the phase 1 -- 2 transition is not a jet break; or GRB~060105 is an outlier of the $E_\gamma$-$E_p$ relation (e.g. \cite{sato06}).
Although the transition without spectral change seems to support the latter, the phase 3 -- 4 transition could correspond to the ``jet-break'' so far recognized in the optical band. In order to examine the relation, in general, detailed investigation is needed in both the early phase X-ray and optical afterglow.
For example, the possible energy injections as we saw in phase 1 and 3 could make deviation from the average fraction of kinetic energy having been estimated only from the prompt gamma-ray emission.

Alternatively, the jet break might have occurred after the observations. 
If we adopt the observed $E_p = 424_{-22}^{+25}$ keV and estimated 
$pz=4.0\pm1.3$ (\cite{gcn4439}) 
with the Ghirlanda relation, the expected jet opening angle $\theta \sim 0.07$ rad and the corresponding jet break time ($\sim 6$ days) is far beyond the end of observation. If that is the case, however, it is very difficult to explain the very steep decay of $\alpha \sim 2.2$.

\subsection{Spectral line feature search}
There are several independent reports on discovery of the prominent emission line features in the X-ray afterglow spectra as mentioned in \S~\ref{sec:intro}. For example, Piro et al. (1999) reported the first detection of iron emission line in the X-ray afterglow spectrum of GRB~970508 though it disappeared in the following flaring activity. Yoshida et al. (2001) found the radiative recombination edge of fully ionized iron in GRB~970828 during a period of flare activity. 
In all these cases, we have no information on the spectrum in the early afterglow until the advent of {\it Swift}.

We performed the line search in the early afterglow of GRB~060105, but found no emission line features, neither in the time averaged spectra from 87--$1.2 \times 10^{5}$ s after the trigger, nor in the time-resolved spectra for the time regions (a) -- (c), including the possible late time refreshed shock phase. 
Thus we have to conclude that this event never exhibited any strong spectral features within the line sensitivity. 
In particular, assuming the redshift of $pz = 4.0$, we obtained extremely low upper limit of iron line equivalent width EW~$\le 15~{\rm eV}$. 
The derived upper limits are very low, an order of magnitude below the claimed iron line detections with {\it Beppo}-SAX (\cite{piro99}), {\it ASCA} (\cite{yoshida01}) and {\it Chandra} (\cite{butler03}). It agrees well with the report from the {\it XMM-Newton} afterglow observations, while although there was a marginal claim of the soft lines, no iron lines were found (e.g. \cite{reeves02}; \cite{watson03}).
On the possibility of the soft lines (e.g. Mg, Si, S), if the pseudo redshift is correct (i.e. $z=4$), then these lines would be shifted below the XIS soft band and would not be detectable. 
Future observations with a low $z$ burst will be required to test for the possibility of soft lines.

\section{Conclusion}
We observed the early X-ray afterglow from the GRB~060105 with {\it Swift} and
{\it Suzaku}. The X-ray lightcurve obtained with {\it Swift}/XRT and 
{\it Suzaku}/XIS exhibited an early steep decay from 0.04 day and lasted for 
0.3 day. After a flattening, the lightcurve shows a steep decay again, and
the slope coincides with what we saw in the first steep decay. 
The X-ray spectral shape, energy index and absorption column are consistent with being constant during the observations, although it infers gradual flattening and steepening through the shallow decay to successive steep decay phase.
The decay indices for the two steep decays suggest a jetlike decay, and phase transitions without spectral changes suggest early jet break, which, however, requires a very narrow jet and the GRB to be an outlier of the Ghirlanda relation. 
We also present a very strict upper limit on the possible atomic spectral lines with the precise spectroscopy with the {\it Suzaku}/XIS and the {\it Swift}/XRT.

\section{Acknowledgement}
Authors thank the {\it Suzaku} operation team, especially to Drs. R. Fujimoto, Y. Terashima, N. Y. Yamasaki, T. Inui, for their support to this quick response observation. Authors also appreciate both the {\it Swift} team and the {\it Suzaku} (ASTRO-E/EII) team member for their devoted work in their development, operation, management and services.
Y.E.N. is supported by JSPS Research Fellowships for Young Scientists.
This research was partially supported by the Ministry of Education, Science, Sports and Culture, Grant-in-Aid for Scientific Research on Priority Areas 14079102-01, 2002 -- 2007.





\begin{thebibliography}{}

\bibitem[Amati et al.(2002)]{amati02}
  Amanti,~L. \etal\ 2002, \aap, 390, 81
\bibitem[Barthelmy et al.(2005)]{barthelmy05}
  Barthelmy,~S.~D., \etal\ 2005, Space Science Reviews, 120, 143
\bibitem[Burrows et al.(2005a)]{burrows05a}
  Burrows,~D.~N., \etal\ 2005a, Space Science Reviews, 120, 165
\bibitem[Burrows et al.(2005b)]{burrows05b}	
  Burrows,~D.~N., \etal\ 2005b, Science, 309, 5742, 1833
\bibitem[Butler et al.(2003)]{butler03}               
   Butler,~N.~R., Marshall,~H.~L., Ricker,~G.~R., Vanderspek,~R.~K., 
   Ford,~P.~G.,
   Crew,~G.~B., Lamb,~D.~Q., \& Jernigan,~J.~G., \ 2003, \apj, 597, 1010
\bibitem[Frail et al.(2006)]{gcn4441}      
   Frail,~D.~A. on behalf of a larger collaboration \ 2006, GCN, 4441
\bibitem[Gehrels et al.(2004)]{gehrels04}
  Gehrels,~N., \etal\ 2004, \apj, 611, 1005
\bibitem[Ghirlanda, Ghisellini \& Lazzati(2004)]{ggl04}
   Ghirlanda,~G., Ghisellini,~G., \& Lazzati,~D. \ 2004, \apj, 616, 331
\bibitem[Godet et al.(2006)]{gcn4433}
   Godet,~O., et al. \ 2006, GCN, 4433
\bibitem[Golenetskii et al.(2006)]{gcn4439}   
   Golenetskii,~S., Aptekar,~ R., Mazets,~E., Pal'shin,~V., Frederiks,~D. 
   Cline,~T. on behalf of the Konus-Wind team \ 2006, GCN, 4439
\bibitem[Izumiura et al.(2006)]{gcn4430}
   Izumiura,~H., Mito, H., Urata,~Y., Huang,~K.~Y., Ip,~W.~H., Qiu,~Y.
   on behalf of EAFON.\ 2006, GCN, 4430
\bibitem[Kalberla et al.(2005)]{nh05}   
   Kalberla,~P.~M.~W., Burton,~W.~B., Hartmann,~D.~A.~E.~M., Bajaja,~E., 
   Morras,~R., \& P\"{o}ppel,~W.~G.~L. \ 2005, \aap, 440, 775 
\bibitem[Kann \& Manohar(2006)]{kann06}
   Kann,~D.~A. \& Manohar,~S.\ 2006, GCN, 5278
\bibitem[Kokubun et al.(2006)]{kokubun06}
  Kokubun,~M., et al. this issue
\bibitem[Koyama et al.(2006)]{xis06}
  Koyama,~K., et al. this issue
\bibitem[Serlemitsos et al.(2006)]{xrt06}
  Selremitsos,~P., et al. this issue
\bibitem[Maeno et al.(2006)]{gcn4431}
   Maeno,~S., Sonoda,~E., Masuda,~S., Nakamura,~Y., Yamauchi,~M.\ 2006, 
   GCN, 4431
\bibitem[Markwardt et al.(2006)]{gcn4435}
   Markwardt,~C., et al.\ 2006 GCN, 4435
\bibitem[Mitsuda et al.(2006b)]{gcn4449}            
   Mitsuda,~K., Fujimoto,~R., Inui,~T., Murakami,~T., \&
   the Suzaku GRB-ToO team \ 2006b, GCN, 4449
\bibitem[Mitsuda et al.(2006a)]{suzaku06}
  Mitsuda,~K., et al.\ 2006a this issue
\bibitem[Nakazawa et al.(2006)]{gcn4606}            
   Nakazawa,~K., \etal\ 2006, GCN, 4606
\bibitem[Nousek et al.(2006)]{nousek06}
   Nousek,~J.~A., \etal\ 2006, \apj, 642, 389
\bibitem[Pelangeon \& J-L. Atteia (2006)]{gcn4442}         
   Pelangeon,~A., \& Atteia,~J-L. \ 2006, GCN, 4442
\bibitem[Ohno et al.(2006)]{ohno06}
  Ohno,~M., et al. \ 2006, Il Nuovo Ciment C, in press
\bibitem[Piran(1999)]{piran99}
   Piran,~R. \ 1999, Physics Reports, 314, 575
\bibitem[Piro et al.(1999)]{piro99}               
   Piro,~L., \etal\ 1999, \apjl, 514, L73
\bibitem[Rees \& M\'{e}sz\'{a}ros(1994)]{rees94}
   Rees,~M.~J., \& M\'{e}sz\'{a}ros,~P. \ 1994, \apjl, 430, L93
\bibitem[Reeves et al.(2002)]{reeves02}
   Reeves,~J.~N., \etal\ 2002, \nat, 416, 463
\bibitem[Rhoads(1999)]{rhoads99}
   Rhoads,~J.~E., 1999, \apj, 525, 737
\bibitem[Roming et al.(2005)]{roming05}
   Roming,~P.~W.~A., \etal\ 2005, Space Science Reviews, 120, 95
\bibitem[Sato et al.(2006)]{sato06}
   Sato,~G.\etal\ 2006, \apj, submitted
\bibitem[Sako et al.(2005)]{sako05}
   Sako,~M., Harrison,~F.~A., \& Rutledge,~R.~E. \ 2005, \apj, 623, 973
\bibitem[Sarapov et al.(2006)]{gcn4454}      
   Sharapov,~D.,  Ibrahimov,~M., Pozanenko,~A., Rumyantsev.~V.
   on behalf of larger GRB follow up collaboration \ 2006, GCN, 4454
\bibitem[Sari \& Piran(1997)]{sari97}
   Sari,~R., \& Piran,~T. \ 1997, \apj, 485, 270 
\bibitem[Sari, Piran \& Halpern(1999)]{sph99}
   Sari,~R., Piran,~T., \& Halpern,~J.~P. \ 1999, \apjl, 519, L17
\bibitem[Schandy et al.(2006)]{gcn4437}
   Schady,~P., Page,~M.,  Cucchiara,~A., Marshall,~F., Gehrels,~N. 
   on behalf of the Swift/UVOT team \ 2006, GCN, 4437
\bibitem[Takahashi et al.(2006)]{takahasi06}
  Takahashi,~T., et al. this issue
\bibitem[Watson et al.(2003)]{watson03}
   Watson,~D., Reeves,~J.~N., Hjorth,~J., Jakobsson,~P., \& Pedersen,~K.
   \ 2003, \apjl, 595, L29
\bibitem[Yamaoka et al.(2006)]{yamaoka06}
  Yamaoka,~K., et al.,\ 2006, SPIE in press
\bibitem[Yanagisawa et al.(2006)]{gcn4436}
   Yanagisawa,~K., Yatsu,~Y., Kawai,~N. on behalf of the Mitsume 
   collaboration \ 2006, GCN, 4436
\bibitem[Yonetoku et al.(2006)]{gcn4438}
   Yonetoku,~D., Murakami,~T., Kodaira,~H., Okuno,~S., Yoshinari,~S.,
   Kidamura,~T., Kobayashi,~Y. on behalf of the Kanazawa team
   \ 2006, GCN, 4438
\bibitem[Yoshida et al.(2001)]{yoshida01}               
   Yoshida,~A., \etal\ 2001, \apjl, 557, L27
\bibitem[Zhang et al.(2006)]{zhang06}               
   Zhang,~B., Fan,~Y.~Z., Dyks, J., Kobayashi, S., M\'{e}sz\'{a}ros,~P., 
   Burrows,~D.~N., Nousek,~J.~A., Gehrels,~N.\ 2006, \apj, 642, 354
\bibitem[Ziaeepour et al.(2006)]{gcn4429}
  Ziaeepour,~H., \etal\ 2006, GCN, 4429
\bibitem[Zimmerman et al.(2006)]{gcn4440}      
   Zimmerman,~N., Tyagi,~S., \& Halpern,~J.~P.\ 2006, GCN, 4440


\end{thebibliography}
\end{document}